\documentclass[a4paper,12pt]{article}
\usepackage[dvips]{graphicx}
\usepackage{array}
\begin{document}
\begin{center}
{\large \bf Fractal properties of the large-scale structure}
\vspace{5ex}

I. K. Rozgacheva$^a$, A. A. Borisov$^b$, A. A. Agapov$^c$,

I. A., Pozdneev$^d$, O. A., Shchetinina$^e$

\vspace{2ex}

Moscow State Pedagogical University, Moscow, Russia

Peoples' Friendship University of Russia, Moscow, Russia
\vspace{1ex}

E-mail: $^a$rozgacheva@yandex.ru, $^b$borisov.andrey03@gmail.com,

$^c$agapov.87@mail.ru, $^d$van@mail.ru, $^e$olesya\_fox87@mail.ru

\vspace{4ex}
\end{center}
\begin{abstract}

The statistical analysis and the spherical wavelet analysis of the SDSS DR7 quasars distribution and of the WMAP CMB anisotropy are performed. We revealed the qualitative agreement between the angular power spectrum of CMB and the angular power spectrum of the quasar distribution on the celestial sphere. The angular correlation function and the angular power spectrum of the quasar distribution may be described by the power laws $\omega\left(\vartheta\right)\sim\vartheta^{1-\gamma}$ and $u_l\sim l^{\gamma_u-3}$ respectively with the averaged parameters $\left\langle\gamma\right\rangle=2{.}08$ and $\left\langle\gamma_u\right\rangle=1{.}92$. The large quasar groups are discovered and they form the fractal set: the relation between their angular size $\vartheta$ and a number of quasar groups $N$ with this size is characterized by a power-law $N\sim\vartheta^{-d}$ with $d\approx 2{.}08$.
\end{abstract}
\vspace{1ex}

\vspace{3ex}

\begin{center}
{\bf 1. Introduction}
\end{center}
\vspace{1ex}

The large-scale structure of the Universe is the structure of the galaxy distribution in whole observable Universe volume. The regions of higher galaxy number density are distinguished in the structure. They are filaments of galaxies and galaxy clusters \cite{ref1}, \cite{ref2} and planes composed of galaxy clusters \cite{ref3}. These planes are also called superclusters. Huge voids are bordered by planes and may be permeated with filaments. Apparently, the large-scale structure is sponge-like \cite{ref4}. Scales of these structures equal tens and hundreds of megaparsecs (Mpc). Investigation of the large-scale structure of the Universe is necessary for understanding of galaxy evolution and physical laws of the Universe evolution.

At present, the geometrical properties of the large-scale structure are investigated through galaxies, galaxy clusters \cite{ref5}, \cite{ref6}, \cite{ref7}, \cite{ref8}, \cite{ref9}, quasars \cite{ref10}, \cite{ref11}, \cite{ref12} and the CMB temperature anisotropy \cite{ref13}, \cite{ref14}.

The procedure of selection of the main markers of the large-scale structure from the general background contains hardly removable mistakes. This is caused by the fact that we cannot eliminate mistakes related to projection of foreground galaxies on a cluster, overlapping of cluster projections and overlapping of small galaxy groups in a narrow cone of view in a two-dimensional projection of the observable galaxy distribution. Besides, there is the selection effect when galaxies with apparent magnitude values close to some average value are included in a cluster (the cluster is supposed to be composed of approximately similar galaxies).

Investigation of three-dimensional galaxy distribution so far gives not quite reliable results due to low quality of sources spectra used for estimation of distances to galaxies through the Hubble law (galaxy velocity is proportional to its distance). In addition to mistakes relating to peculiar (non-Hubble) galaxy motion, estimation of distance through redshift depends on choice of cosmological model in which the distance is computed as a function of redshift (redshift-space distortion).

Therefore, study of statistical and topological properties of the large-scale structure through only two-dimensional galaxy distribution in large areas of the celestial sphere is relevant. Of course, two-dimensional projection distorts quantitative properties of the large-scale structure. However, two-dimensional projection enables to search for the universal properties of the galaxy distribution which is typical for the whole observable Universe due to high precision of the galaxy equatorial coordinates estimation and large areas of modern surveys. The main difficulty of such researches is related to the fact that wide-band surveys are composed of narrow-band surveys which differ in precision of coordinates estimation and sensitivity of radiation flux measurement. This lead to inhomogeneity of resulting survey. Processing of such inhomogeneous survey may lead to revealing of fictitious voids and clumps.

In this paper the processing of available at present data is performed:

- on the quasar distribution on the celestial sphere (according to the seventh data release of the SDSS quasar catalogue \cite{ref15});

- on the CMB temperature anisotropy \cite{ref16}.

The purpose of the analysis is finding of the most general properties of the large-scale structure. The main result of the work is revealing of the fractal properties of the quasar distribution on the celestial sphere.
\vspace{3ex}

\begin{center}
{\bf 2. Photometrical properties of the SDSS quasars}
\end{center}
\vspace{1ex}

The SDSS catalogue is composed of eight surveys of the celestial sphere. In the area of the celestial sphere with equatorial coordinates $9^h<\alpha <16^h$, $0^{\circ}<\delta <55^{\circ}$ the catalogue contains 105,783 quasars with redshifts $0{.}0645\le z\le 5{.}4608$.

Areas of these eight surveys overlap. On the one hand, this permits to improve the precision of the coordinate estimation in overlapping regions: the survey-averaged errors of source's coordinates estimation equal $\Delta\alpha=0{.}139^{\prime\prime}$, $\Delta\delta =0{.}130^{\prime\prime}$. On the other hand, one can detect more sources in the overlapping regions. Therefore, the number density of detected quasars is higher in these areas. So far one cannot find out if these source clumps are real or related to inhomogeneity of the SDSS catalogue only.

We performed the SDSS-quasars distribution analysis in framework of standard cosmological model with parameters: $H_0=70\ km\cdot s^{-1}\cdot Mpc^{-1}$ is the Hubble constant, $\Omega_M=0{.}3,\ \Omega_{\Lambda}=0{.}7$ are the dimensionless density parameters of dust and $\Lambda$-term respectively. Cosmological distance to a quasar with redshift $z$ (comoving distance) is determined by the formula:
$$
r=\frac{c}{H_0}\int\limits_0^z\frac{dz^{\prime}}{\sqrt{\Omega_M\left(1+z^{\prime}\right)^3+\Omega_{\Lambda}}}.
$$
Distances measured in the Hubble distance $c/H_0$ are used below.

Quasar distribution in right ascension (the equatorial coordinate $\alpha$) and cosmological distance $rH_0/c$ is shown on fig.~1. As one can see, there was an epoch of high galaxy activity in the Universe evolution. One of the fundamental problems of quasar physics is why this epoch was and why there are no quasars in the vicinity of the Local Group of galaxies.

\begin{figure}[h]
\begin{center}
\includegraphics[scale=1]{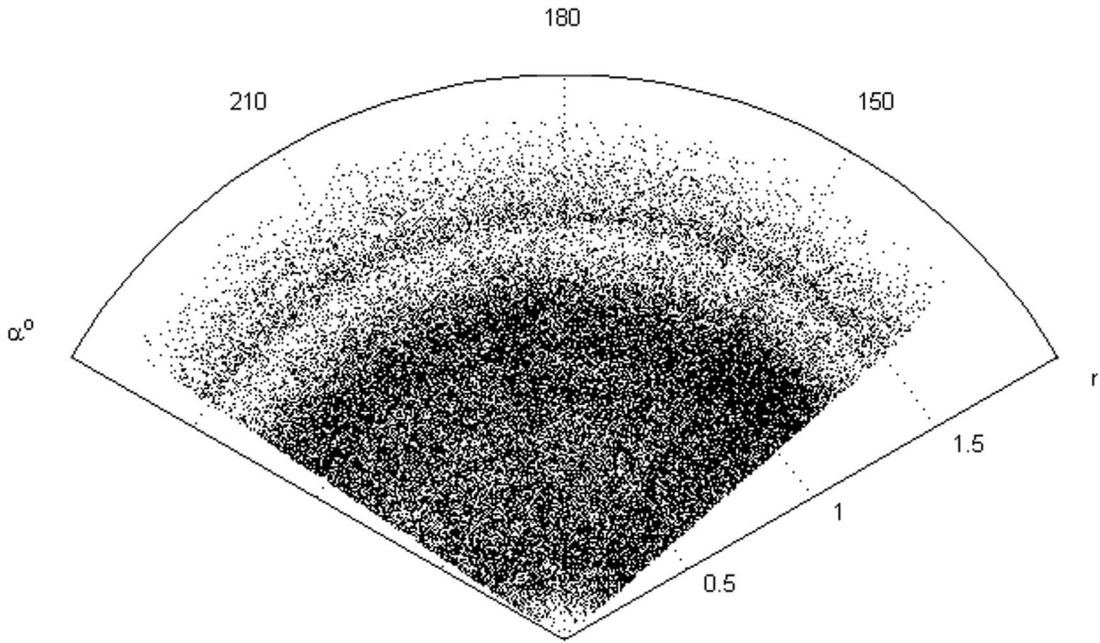}
\caption{\small The quasar distribution: cosmological distance - right ascension}
\end{center}
\end{figure}

Variation of SDSS-quasars number density in an element of comoving volume of spatial cone of the SDSS catalogue is shown on fig.~2 (the number density is normalized to a maximal value). The region of higher number density corresponding to the redshift range $0{.}35<z<2{.}30$ is distinguished. This range is singled out by the catalogue authors after taking into account of constant biases in quasar redshift measurements.

Existence of high galaxy activity epoch indicates that quasars are generally not "standard candles" through which the large-scale structure should be studied. However, astronomers have no other so bright sources. Therefore, we used the five-color photometric system of the SDSS catalogue \cite{ref17} for estimation of SDSS-quasars luminosity function.

\begin{figure}[t]
\begin{center}
\includegraphics[scale=0.8]{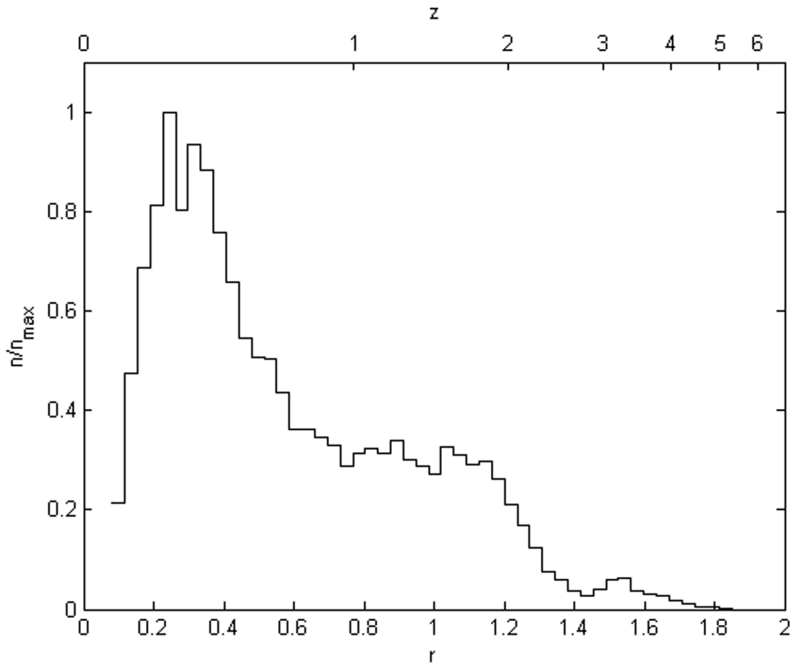}
\caption{\small Variation of SDSS-quasars number density in an element of comoving volume}
\end{center}
\end{figure}

Mean wavelength values of different bands of the system are shown in Table 1.

\begin{center}
\begin{table}[b]
\caption{\small Mean wavelength values of different bands of the SDSS photometric system}
\vspace{2ex}
\begin{center}
\tabcolsep=0.25cm
\extrarowheight=1ex
\begin{tabular}{|c||c|c|c|c|c|}
\hline
Wavelength band               & $u_m$ & $g_m$ & $r_m$ & $i_m$ & $z_m$ \\[6pt]
\hline
Mean value $\left(\AA\right)$ & 3551 & 4686 & 6165 & 7481 & 8931 \\[6pt]
\hline
\end{tabular}
\end{center}
\end{table}
\end{center}
\vspace{1ex}

We have plotted two-dimensional diagrams for different pairs of wavelength bands. It has emerged that there is a relatively regular relation only between apparent magnitudes in a long-wave part of the photometrical system $i_m$ and $z_m$. The diagram $i_m-z_m$ is shown on fig.~3.

\begin{figure}[h]
\begin{center}
\includegraphics[scale=0.6]{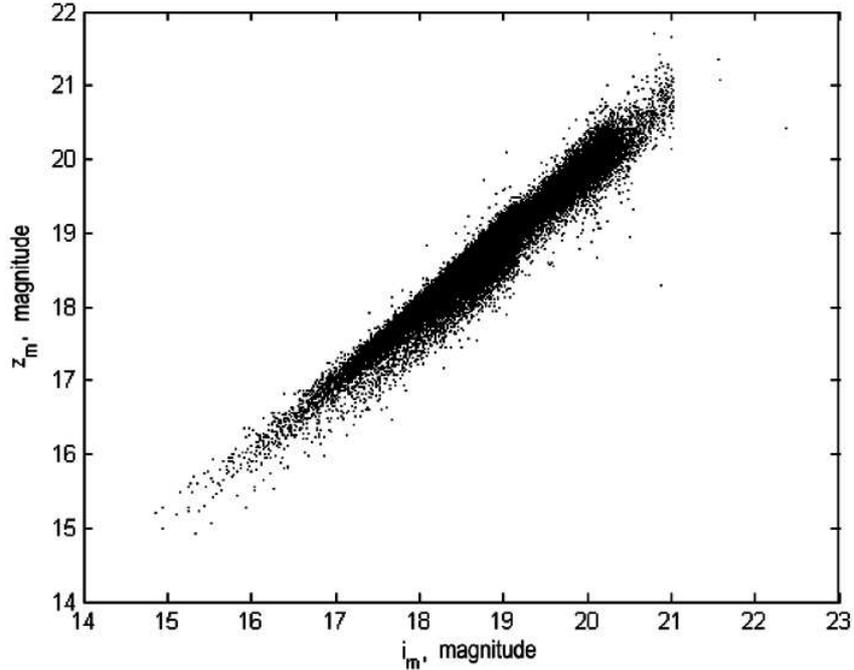}
\caption{\small The diagram of $i_m-z_m$ relation for SDSS quasars}
\end{center}
\end{figure}

The regular relation between $i_m$ and $z_m$ magnitudes holds out a hope that the quasar luminosity function obtained for these bands is biased by observation errors in a less degree.

For decrease of significance of evolution effects we have divided the epoch of the redshift range $0{.}35<z<2{.}30$ into six layers. The division is performed according to two criteria: 1) every layer contains comparable quasar numbers; 2) the quasar luminosity function plots for different layers (for $i_m$ and $z_m$ magnitudes) are qualitatively similar. The second criterion follows from the following reasoning. The luminosity function describes the luminosity distribution of quasars. According to the adopted at present hypothesis, the higher quasar luminosity the higher its mass. Therefore, the luminosity function describes the mass distribution of quasars. If the mass distributions of quasars are similar in every layer we may expect that the same common properties of the large-scale structure are displayed in every layer.

The luminosity function equals number density of quasars with absolute magnitudes $\left(M_i,\ M_i+\Delta M_i\right)$ (for wavelength band $i$) in a comoving volume of a layer:
$$
\varphi_I=\frac{N\left(M_i,\ M_i+\Delta M_i\right)}{V\left(r\right)\Delta M_i}. \eqno(1)
$$
The quasar absolute magnitudes are adduced in the data base \cite{ref17}.

Direct computation by the formula (1) shows that the quasar luminosity function plots for chosen layers are qualitatively similar in spite of unavoidable luminosity selection (among the distant sources we can see only the brightest ones).

\begin{figure}[h!]
\begin{center}
\includegraphics[scale=0.58]{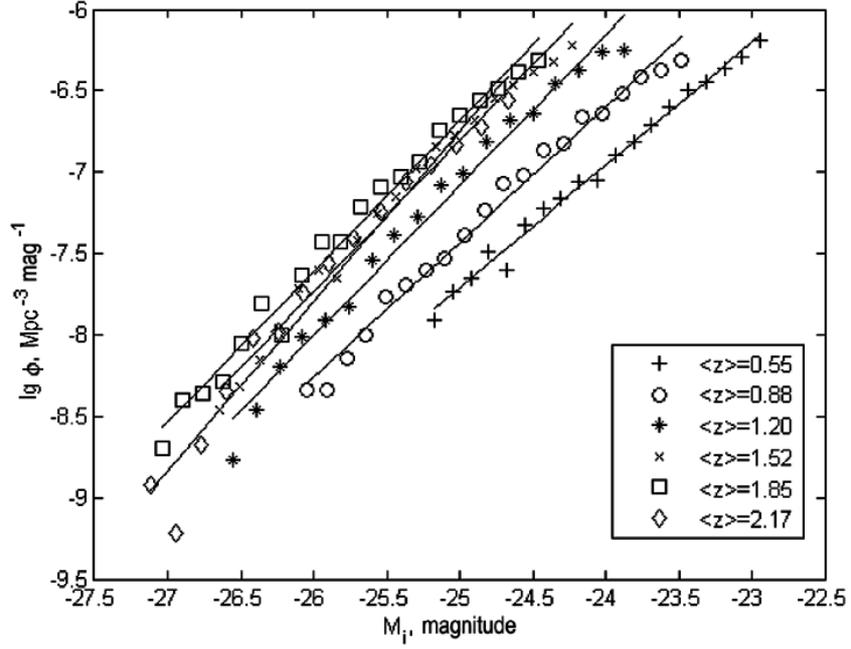}
\caption{\small The quasar luminosity function in different redshift layers for $i_m$ filter. The mean redshift values for every layer are shown in the inner table}
\end{center}
\end{figure}

\begin{figure}[h!]
\begin{center}
\includegraphics[scale=0.58]{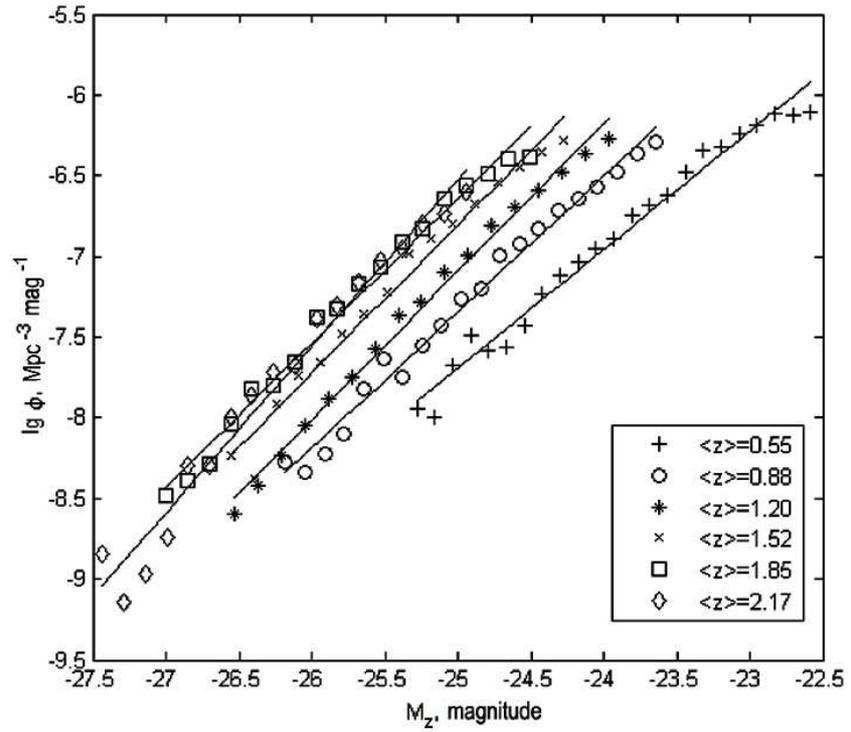}
\caption{\small The same as on fig. 4 for $z_m$ filter}
\end{center}
\end{figure}

The SDSS-quasars luminosity functions for $i_m$ and $z_m$ filters are shown on fig.~4 and 5 respectively. The empirical quasar luminosity function is usually approximated by a two power-law function \cite{ref18}:
$$
\phi\left(M\right)=\phi^*\cdot\left\lbrack 10^{0{.}4\left(\alpha+1\right)\left(M-M^*\right)}+10^{0{.}4\left(\beta+1\right)\left(M-M^*\right)}\right\rbrack^{-1},
$$
where $\alpha$ and $\beta$ relate to branches of bright sources faint sources respectively, $M^*$ is a characteristic magnitude value. The SDSS-catalogue data don't permit to establish the luminosity function shape for faint quasars due to selection effects. Therefore, we determine the parameter $\alpha$ of the empirical power-law luminosity function for the bright sources branch: $\phi\left(M\right)=\phi^*\cdot10^{-0{.}4\left(\alpha+1\right)M}$.

The parameters of the redshift layers and of the quasar empirical power-law luminosity function for $i_m$ and $z_m$ wavelength filters are given in Table 2.

\begin{center}
\begin{table}
\caption{\small The parameters of the redshift layers and of the quasar empirical power-law luminosity function for $i_m$ and $z_m$ wavelength filters}
\vspace{1ex}
\begin{center}
\tabcolsep=0.23cm
\extrarowheight=1ex
\begin{tabular}{|c||c|c|c|c|c|}
\hline
Redshift layer & $\left\langle z\right\rangle$ & Layer bounds & Quasar number & $\alpha$ ($i_m$ filter) & $\alpha$\ ($z_m$ filter) \\[6pt]
\hline
\hline
1 & 0{.}55 & $0{.}39<z<0{.}71$ & 6471 & -2{.}88 & -2{.}83 \\[6pt]
\hline
2 & 0{.}88 & $0{.}71<z<1{.}04$ & 7145 & -3{.}07 & -3{.}11 \\[6pt]
\hline
3 & 1{.}20 & $1{.}04<z<1{.}36$ & 9065 & -3{.}31 & -3{.}30 \\[6pt]
\hline
4 & 1{.}52 & $1{.}36<z<1{.}68$ & 10060 & -3{.}33 & -3{.}30 \\[6pt]
\hline
5 & 1{.}85 & $1{.}68<z<2{.}01$ & 9420 & -3{.}32 & -3{.}24 \\[6pt]
\hline
6 & 2{.}17 & $2{.}01<z<2{.}33$ & 4982 & -3{.}60 & -3{.}58 \\[6pt]
\hline
\end{tabular}
\end{center}
\end{table}
\end{center}
\vspace{1ex}

Figures 4 and 5 shows that the luminosity function curves for layers 4, 5 and 6 intersect. This may be related to either observation errors or quasars physical evolution. For example, galaxy mergers occur in regions of higher galaxy number density. This stimulate galaxy cores activity and appearance of quasars. Due to this, the mass distribution of galaxies changes and the luminosity function shape changes as well.
\vspace{3ex}

\begin{center}
{\bf 3. Angular correlation function and power spectrum of SDSS quasars}
\end{center}
\vspace{1ex}

For general description of the quasar distribution we have computed correlation dimension, angular two-point correlation function and angular power spectrum for the expansion of the observable quasar distribution in spherical functions in each redshift layer.

The correlation dimension computation methods are given in papers \cite{ref19}, \cite{ref20}, \cite{ref21}, for example.

The correlation dimension characterizes quasar clumping degree and difference of the quasar distribution from a homogenous and isotropic one. The dependence of quasar number $N\left(r\right)$ in a sphere on its radius $r$ for the chosen redshift range $0{.}39<z<2{.}33$ is shown on fig.~6. The points distribution on the plot may be described with a power-law with a high correlation coefficient value ($0{.}999$):
$$
N\left(<r\right)\sim r^{d_c}, \eqno(2)
$$
where the exponent (correlation dimension) equals $d_c=2{.}17$.

The power-law (2) is usually considered as an indication of fractality of a spatial sources distribution. Similar $N\left(r\right)$ dependence is typical for galaxies $\left(d_c\approx 1{.}15\div 2{.}25\right)$ and it is a large-scale structure common law \cite{ref20}, \cite{ref21}.

\begin{figure}[t]
\begin{center}
\includegraphics[scale=0.8]{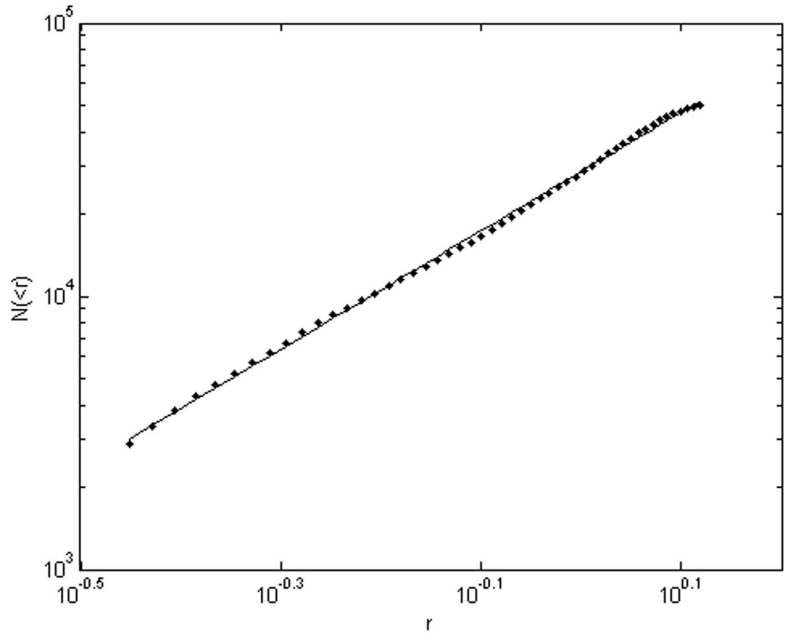}
\caption{\small $N-r$ relation for $0{.}39<z<2{.}33$}
\end{center}
\end{figure}

For each layer we have computed the quasar angular correlation function $\omega\left(\vartheta\right)$ through the usually used formula:
$$
\omega\left(\vartheta\right)=\left(\frac{N_p}N\right)\frac{m\left(\vartheta\right)}{m_p\left(\vartheta\right)}-1, \eqno(3)
$$
where $\vartheta$ is measured angular distance between quasars, $N$ is a quasar number in a sample, $N_p$ is point number in random Poisson distribution in a celestial sphere area of the SDSS catalogue, $m\left(\vartheta\right)$ is a number of quasar pairs with mutual distance $\vartheta$, $m_p\left(\vartheta\right)$   is a point pairs number in random Poisson distribution with the same distance $\vartheta$. angular distance on the celestial sphere using in expression (3) is calculated through the formula:
$$
\vartheta=\arccos{\left(\cos{\delta_1}\cos{\delta_2}\cos{\left(\alpha_1-\alpha_2\right)}+\sin{\delta_1}\sin{\delta_2}\right)},
$$
where $\delta_1$, $\delta_2$, $\alpha_1$ and $\alpha_2$ are declinations and right ascensions of two quasars.

Every layer presents a large-scale structure variant in an epoch corresponding to a mean redshift of the layer. If the cosmological principle (statistical homogeneity and isotropy of the Universe) is satisfied the correlation functions of different layers must be similar irrespective of errors related to overlapping of two-dimensional projections of quasar clumps and catalogue inhomogeneity.

The correlation function shape is revealed to be similar for each of six layers. An example of correlation function plot for the layer $1{.}04<z<1{.}36$ is shown on fig.~7.

\begin{figure}[h]
\begin{center}
\includegraphics[scale=0.8]{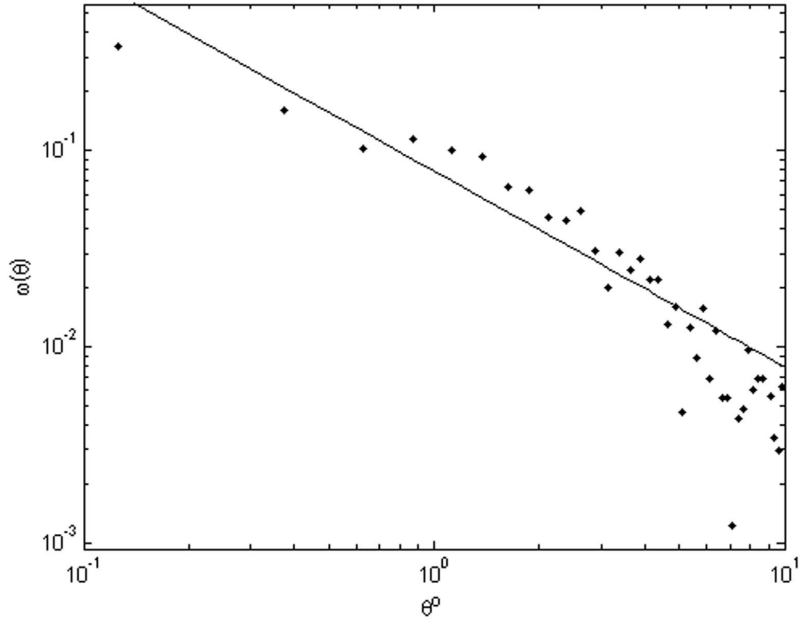}
\caption{\small The correlation function of SDSS quasars for the layer $1{.}04<z<1{.}36$}
\end{center}
\end{figure}

We have computed angular power spectrum of SDSS quasars for each layer using methods described in \cite{ref23}, \cite{ref24}. The power spectrum is usually used for comparison of large-scale structure formation models with observations because the power spectrum depends on an adopted cosmological model. This fact enables to estimate cosmological parameters.

We describe the quasar distribution on the celestial sphere by a sum of $\delta$-functions:
$$
f\left(\vartheta,\varphi\right)=\frac1{\sin{\vartheta}}\sum_{k=1}^N\delta\left(\vartheta-\vartheta_k\right)\delta\left(\varphi-\varphi_k\right)=\sum_{l=0}^{\infty}\sum_{m=-l}^la_l^mY_l^m. \eqno(4)
$$
The coefficients $a_l^m$ of the expansion in spherical functions $Y_l^m$ are determined by the formula:
$$
a_l^m=\sum_{k=1}^NY_l^{m^*}\left(\vartheta_k,\varphi_k\right). \eqno(5)
$$

For a catalogue covering only a part of the celestial sphere, estimations $b_l^m$ of true expansion coefficient values are calculated through the expression:
$$
b_l^m=\frac{\displaystyle\left|a_l^m-I_l^mN/\Omega\right|^2}{\displaystyle J_l^m}, \eqno(6)
$$
where the integrals $\displaystyle I_l^m=\int\limits_{\Omega}{}Y_l^md\Omega$, $\displaystyle J_l^m=\int\limits_{\Omega}{}\left|Y_l^m\right|^2d\Omega$. For the whole sphere $I_l^m=0$, $J_l^m=1$. The integrals $I_l^m$ and $J_l^m$ are computed numerically.

The angular power spectrum estimation $u_l$ is given by the formula:
$$
u_l=\frac{\left(b_l^m\right)_m}{N/\Omega}-1, \eqno(7)
$$
where
$$
\left(b_l^m\right)_m=\frac{\displaystyle \sum_{m=-l}^lb_l^mJ_l^m}{\displaystyle \sum_{m=-l}^lJ_l^m}
$$
is weight-average of coefficients (6) with weights $J_l^m$.

An example of SDSS quasars angular power spectrum (7) plot for the layer $1{.}04<z<1{.}36$ is shown on fig.~8. The angular power spectrum shapes for all six layers are similar.

\begin{figure}[h]
\begin{center}
\includegraphics[scale=0.8]{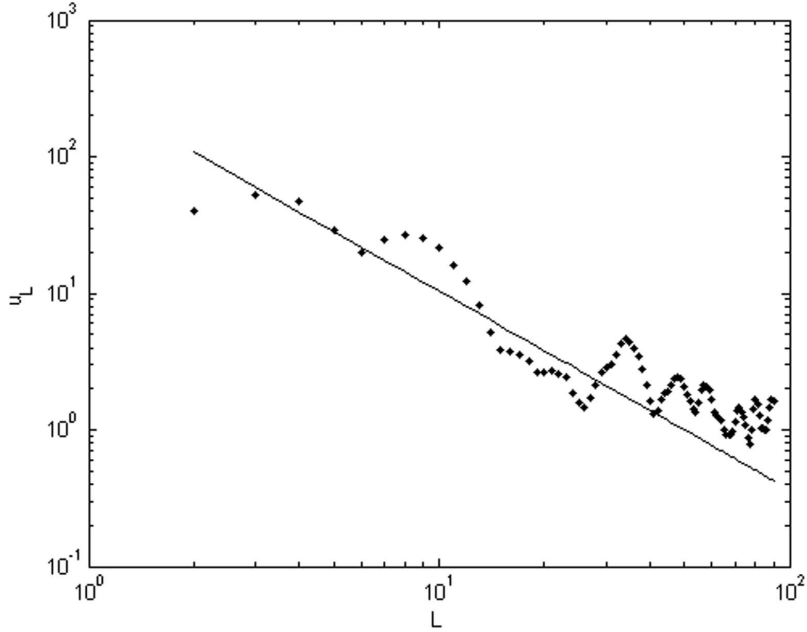}
\caption{\small Angular power spectrum of SDSS quasars in the layer $1{.}04<z<1{.}36$}
\end{center}
\end{figure}

In case of power-law (2), the angular correlation function and the power spectrum are power-laws as well:
$$
\omega\left(\vartheta\right)\sim\vartheta^{1-\gamma}, \eqno(8)
$$
$$
u_l\sim l^{\gamma_u-3}. \eqno(9)
$$
The least-squares lines plotted on figures 7 and 8 illustrate the power-laws (8) and (9).

The parameter values of laws (8) and (9) for six layers are given in table 3. The first column is layer's number, the second column is the exponent $\gamma$, the third one is correlation coefficient values for correlation function plot points (fig.~7), the fourth one is the exponent $\gamma_u$, the fifth column is correlation coefficient values for power spectrum plot points (fig.~8).

The layer-averaged exponent values of laws (8) and (9) are close: $\left<\gamma\right>=2{.}08$, $\left<\gamma_u\right>=1{.}92$.

The power laws (2), (8) and (9) indicate possibility of fractal properties of the large-scale structure \cite{ref20}. Note that when these power-laws are satisfied exactly in regions of high quasar number density (i.e. $\omega\left(\vartheta\right)>1$), the equations $d_c=3-\gamma$ and $\gamma=\gamma_u$ are satisfied \cite{ref24}. The SDSS-quasars correlation function (3) is less than unity, therefore, these equations are not satisfied for obtained values of the correlation dimension $d_c=2{.}17$ and parameters $\gamma$ and $\gamma_u$.

\begin{center}
\begin{table}
\caption{\small Correlation function and power spectrum parameters for six redshift layers}
\vspace{1ex}
\begin{center}
\tabcolsep=0.25cm
\extrarowheight=1ex
\begin{tabular}{|c||c|c|c|c|}
\hline
Layer \# & $\gamma$ & $r_{\omega}$ & $\gamma_u$ & $r_u$ \\[6pt]
\hline
\hline
1 & 2{.}02 & -0{.}71 & 2{.}06 & -0{.}77 \\[6pt]
\hline
2 & 2{.}23 & -0{.}85 & 2{.}00 & -0{.}94 \\[6pt]
\hline
3 & 2{.}25 & -0{.}87 & 1{.}84 & -0{.}90 \\[6pt]
\hline
4 & 2{.}00 & -0{.}92 & 1{.}87 & -0{.}92 \\[6pt]
\hline
5 & 1{.}86 & -0{.}94 & 1{.}84 & -0{.}92 \\[6pt]
\hline
6 & 2{.}09 & -0{.}90 & 1{.}91 & -0{.}92 \\[6pt]
\hline
\end{tabular}
\end{center}
\end{table}
\end{center}
%\vspace{3ex}

\begin{center}
{\bf 4. Fractal properties of the CMB temperature anisotropy according to WMAP data}
\end{center}
\vspace{1ex}

CMB photons show us the Universe as it was at the recombination epoch. The WMAP experiment proves convincingly that CMB temperature angular fluctuations exist.

For analysis of the CMB temperature anisotropy in WMAP experiment the expansion of two-dimensional temperature field in spherical functions is used:
$$
T\left(\vartheta,\varphi\right)=\sum_{l=0}^{\infty}\sum_{m=-l}^la_l^mY_l^m,
$$
where $a_0^0=T_0$ is the mean temperature value.

An integer number of each spherical harmonic $Y_l^m$ fits in the sphere. A multipole number $l$ shows how many periods of the harmonic fits in the interval $\left[0,\pi\right]$. A positive part of a harmonic corresponds to a patch with positive temperature. The patch size is of the order of $\displaystyle\theta\approx\frac{\pi}l=\frac{180^{\circ}}l$.

The quantity $\tilde C_l$ defined through the expression
$$
\left<a_{lm}a_{l^{\prime}m^{\prime}}^*\right>=\tilde C_l\delta_{ll^{\prime}}\delta_{mm^{\prime}}
$$
is the power spectrum. The angular brackets denote averaging over several scannings of the celestial sphere. The power spectrum permits to single out the CMB temperature anisotropy for different angular scales.

The power spectrum averaged over all $m$ modes (their number is $2l+1$) is used in investigations:
$$
C_l=\frac1{2l+1}\sum_{m=-l}^l\left<a_{lm}a_{lm}^*\right>.
$$

Generally, power spectrum values depend on observer's location in the Universe. For statistical description of the CMB temperature anisotropy a value averaged over all observers (realizations of observations) is desirable. We have one realization only; therefore, for the WMAP data processing the "gaussianity hypothesis" is adopted: temperature fluctuations are independent and their amplitudes satisfy the Gaussian statistics for all observers. In this case, a relative error of the power spectrum is \cite{ref25}
$$
\frac{\Delta\tilde C_l}{\tilde C_l}=\sqrt{\frac2{2l+1}}.
$$

The power spectrum dependence on values of mean multipoles is shown on fig.~9. The least-squares line is described with a power-law:
$$
C_l\sim l^{-1{.}74}. \eqno(10)
$$
Correlation coefficient equals $0{.}96$. This high value implies that the dependence (10) is a significant statistical correlation.

\begin{figure}[h]
\begin{center}
\includegraphics[scale=0.8]{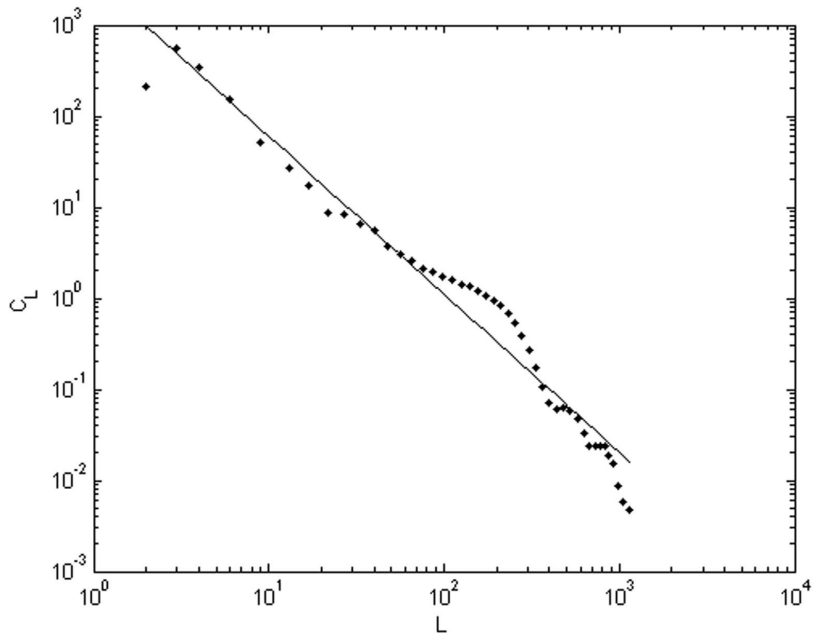}
\caption{\small CMB power spectrum}
\end{center}
\end{figure}

The power spectrum of the CMB temperature fluctuations qualitatively corresponds to the power spectrum of the quasar distribution (9): $u_l\sim l^{-1{.}08}$.

Peaks of CMB power spectrum plot are more relief if the function $\displaystyle\frac{l\left(l+1\right)}{2\pi}C_l$ is used.

What the fractal properties of the large-scale structure (2), (8), (9) and (10) are indicative of? The obvious answer is that the properties are consequences of the fractal properties of the initial matter density perturbations which further led to star, galaxy and cluster formation due to gravitational instability. Now we observe traces of these fractal properties through quasars. This interpretation follows from the hypothesis of Gaussian (thermal) spectrum of the initial density perturbations in Newtonian approximation \cite{ref26}. However, it is not quite so in the general theory of relativity because Einstein's tensor is not invariant under scale transformation of the Riemannian space-time \cite{ref27}, \cite{ref28}. If the large-scale structure evolution is described by Einstein's gravity theory the fractal properties may not conserve, even if the initial fluctuations had the thermal spectrum. In paper \cite{ref26} a cosmological model is suggested in framework of which the observable fractal properties may be explained. It has emerged that fractality may follow from the fact that matter energies in clumps don't change and are related to each other by a power transformation. The simplest consequence of this condition is a case when clumps sizes form a geometric progression. In the next section we show that SDSS-quasar clumps satisfy this condition.
\vspace{3ex}

\begin{center}
{\bf 5. Wavelet analysis of the quasar distribution on the celestial sphere}
\end{center}
\vspace{1ex}

For estimation of quasar clump sizes we used the spherical wavelet transform \cite{ref29}.

A spherical wavelet is a function $\psi\in L^2\left(S^2\right)$ defined on a sphere and satisfying a zero mean condition:
$$
C_{\psi}\equiv\int\limits_{S^2}^{}\frac{\psi\left(\vartheta,\varphi\right)}{1+\cos{\vartheta}}\sin{\vartheta}d\vartheta d\varphi=0.
$$
The spherical wavelet $\psi$ is associated with a flat wavelet $\psi_E$ (in polar coordinate system) through the inverse stereographic projection:
$$
\psi\left(\vartheta,\varphi\right)=\frac{2\psi_E\left(2\tan{\displaystyle\frac{\vartheta}2},\varphi\right)}{1+\cos{\vartheta}}.
$$
If the flat wavelet is the "mexican hat"
$$
\psi_E\left(\rho,\varphi\right)=\frac1{\sqrt{2\pi}}\left(2-\rho^2\right) e^{-\rho^2/2},
$$
the spherical wavelet equals
$$
\psi\left(\vartheta,\varphi\right)=\frac4{\sqrt{2\pi}}\frac{\left(1-2\tan^2{\displaystyle\frac{\vartheta}2}\right)}{1+\cos{\vartheta}}\exp{\left(-2\tan^2{\frac{\vartheta}2}\right)}. \eqno(11)
$$
The spherical wavelet (11) is an axisymmetric function.

Scaling of the spherical wavelet is performed through the formula
$$
\psi_{a,0,0}\left(\vartheta,\varphi\right)=\frac{2a}{\left(a^2-1\right)\cos{\vartheta}+\left(a^2+1\right)}\psi\left(\vartheta_a\right),
$$
where
$$
\tan{\frac{\vartheta_a}2}=\frac1a\tan{\frac{\vartheta}2}.
$$
A displacement is performed through the replacement of the angle $\vartheta$ by the angle
$$
\vartheta'=\arccos{\left(\cos{\vartheta}\cos{\alpha}+\sin{\vartheta}\sin{\alpha}\cos{\left(\varphi-\beta\right)}\right)},
$$
$\alpha$ and $\beta$ are the parameters of the displacement on the sphere.

The spherical wavelet transform of function $f\left(\vartheta,\varphi\right)$ is defined by the formula
$$
W\left(a,\alpha,\beta\right)=\int\limits_{S^2}^{}\overline{\psi_{a,\alpha,\beta}\left(\vartheta,\varphi\right)}f\left(\vartheta,\varphi\right)\sin{\vartheta}\ d\vartheta\ d\varphi, \eqno(12)
$$
where the bar indicates complex conjugation. The integral (12) is composed of positive and negative parts. The positive part arises if there are points under the hat. The negative part arises if there are points under the hat brim. Therefore, the wavelet coefficient (12) is positive if point number density under the hat is more than number density under the hat brim. The coefficient is negative if number density under the brim is more than that under the hat. If number density is equal everywhere the wavelet coefficient (12) is zero.

Thus, wavelet coefficients (12) describe gradients of surface number density of the quasar distribution. The wavelet analysis permits to single out quasar clump areas: the wavelet coefficients (12) are positive in these areas. Sizes of the areas are of the order of the "mexican hat" size $a$. The more the "mexican hat" size the lower the wavelet analysis resolution, because small clump areas and rarefaction areas of the quasar distribution get under the hat when size $a$ is large.

Estimation of true clump sizes is a difficult problem because the wavelet coefficient value (12) depends on a clump size, on a number of quasars in the clump and on a quasar distribution in the clump. Therefore, there is no definite relation between the size and the wavelet coefficient value (12) of the clump.

For solution of this problem we should ascertain a relation between angular size $\vartheta$ of a clump and size of the "mexican hat" $a$. The power spectrum $C_l\left(a\right)$ for the spherical "mexican hat" (11):
$$
c_l^m\left(a\right)=\int\limits_{4\pi}^{}\psi_{a,0,0}Y_l^md\Omega,
$$
$$
C_l\left(a\right)=\frac1{2l+1}\sum_{m=-l}^l\left|c_l^m\left(a\right)\right|^2.
$$

This power spectrum $C_l\left(a\right)$ as a function of multipole moment has one peak at value $l_{max}$. Therefore, the spherical wavelet (11) singles out clumps with angular radius ${\displaystyle\frac{180^{\circ}}{l_{max}}\approx\vartheta_a}$. One can find out numerically that the angular size related to the size of "mexican hat" through expression: ${\displaystyle a\frac{180^{\circ}}{\sqrt{\pi}}\approx\frac{180^{\circ}}{l_{max}}\approx\vartheta_a}$.

Thus, the continuous wavelet transform (12) in which the spherical "mexican hat" (11) is used as a analyzing wavelet has resolution of $\sqrt{\pi}a$ radian for every scaling coefficient $a$.

We describe the distribution of $N$ quasars on the celestial sphere through a sum of $\delta$-functions (4). The wavelet transform of the function (4) equals
$$
W\left(a,\alpha,\beta\right)=\sum_{i=1}^N\overline{\psi_{a,\alpha,\beta}\left(\vartheta_i,\varphi_i\right)}=\left(2/\pi\right)^{1/2}\frac1a\sum_{i=1}^N\left(1+x_i^2\right)^2\frac{a^2-2x_i^2}{a^2+x_i^2}\exp{\left(-2\frac{x_i^2}{a^2}\right)}, \eqno(13)
$$
where
$$
x_i=\tan{\frac{\tilde\vartheta_i}2}=\sqrt{\frac{1-\cos{\tilde\vartheta_i}}{1+\cos{\tilde\vartheta_i}}}=\sqrt{\frac{1-\cos{\vartheta_i}\cos{\alpha}-\sin{\vartheta_i}\sin{\alpha}\cos{\left(\varphi_i-\beta\right)}}{1+\cos{\vartheta_i}\cos{\alpha}+\sin{\vartheta_i}\sin{\alpha}\cos{\left(\varphi_i-\beta\right)}}}.
$$

The map of distribution of the wavelet coefficient (13) values for $a=0{.}008$ radian is shown on fig.~10. Quasar clumps (large quasar groups) look like light spots. The more quasar number in a clump the higher brightness of a spot (for spots with equal size).

\begin{figure}[h]
\begin{center}
\includegraphics[scale=1]{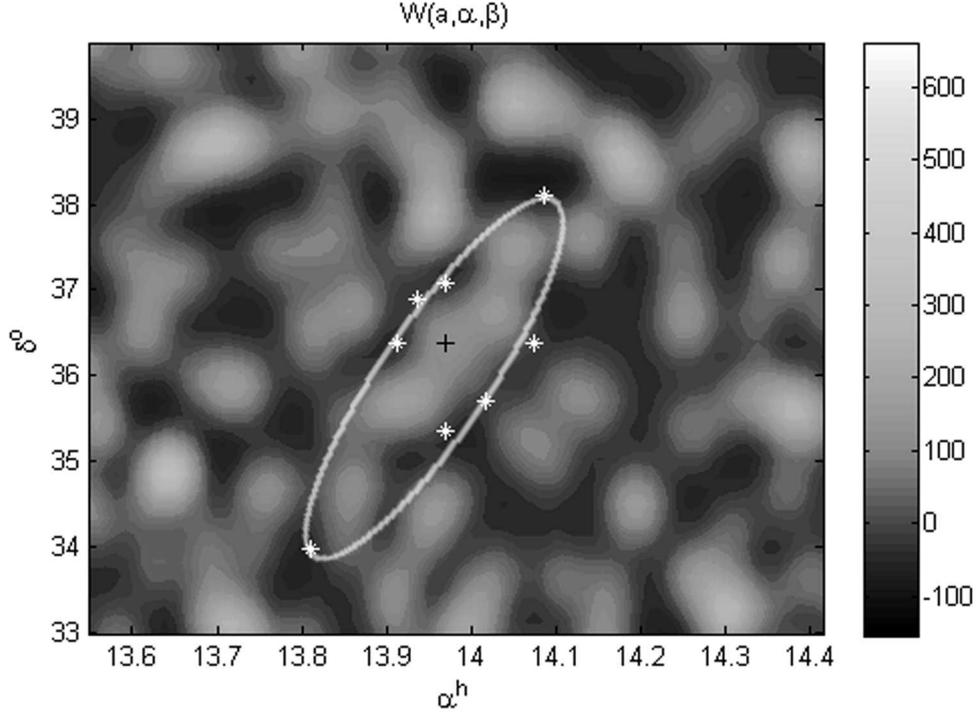}
\caption{\small A fragment of the wavelet coefficient (13) value distribution and an example of a selection of a large quasar chain}
\end{center}
\end{figure}

For study of fractal properties of large quasar groups we should determine a number of groups $N_c$ with a certain angular size $\vartheta_c$. For this, we need to adopt a criterion for determination of group size. As noted above, a wavelet coefficient characterizes a gradient of quasar distribution number density. If the wavelet coefficient is negative, there is a region of lower quasar number density (rarefaction) in that place. Estimation of quasar group sizes is carried out as follows. We determine positions of local peaks of the wavelet coefficients. These peaks usually group in patches of positive wavelet coefficient values: one patch may contain from one to some tens of such peaks. A bound of every patch is approximated by an ellipse (fig.~10) a major semiaxis of which is considered as a group size. For every angular size value $\vartheta_c$ one can compute a number of equal large quasar groups $N_c$.

The dependence $N_c\left(\vartheta_c\right)$ for the layer $1{.}04<z<1{.}36$ is shown on fig.~11. This is well described by a power law (with correlation coefficient value -0,98):
$$
N_c\sim\vartheta_c^{-2{.}08}. \eqno(14)
$$

\begin{figure}[h]
\begin{center}
\includegraphics[scale=0.8]{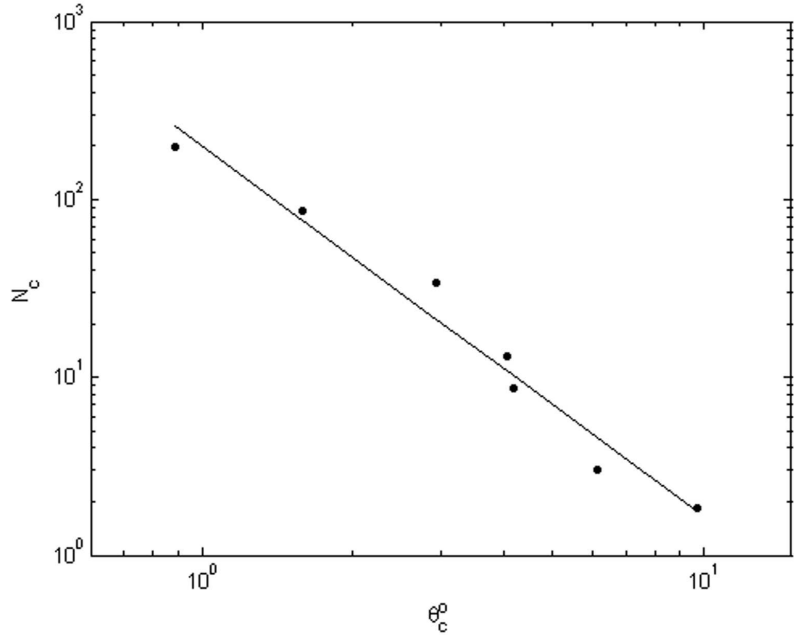}
\caption{\small $N_c\sim\vartheta_c^{-d_c}$ relation for the layer $1{.}04<z<1{.}36$}
\end{center}
\end{figure}

Thus, the large quasar group distribution in angular sizes is the power law (14). This distribution indicates that the large-scale is a fractal set and the exponent $d\approx 2{.}08$ is its fractal dimension \cite{ref30}. The power law may be a consequence of the fact that large quasar groups are self-similar. For example, let their angular sizes form a geometric progression
$$
\vartheta_i=\vartheta_0q^{i-1}, \eqno(15)
$$
where $q<1$ is a geometric ratio. Then, a number of quasar groups with sizes $\vartheta_i\ge\vartheta_n=\vartheta_0q^{n-1}$ satisfies the law (14):
$$
N_n=\sum_{i=1}^nN_i\sim\sum_{i=1}^n\vartheta_i^{-d}=\vartheta_0^{-d}\frac{q^{-dn}-1}{q^{-d}-1}\approx\left(\vartheta_0q^{n-1}\right)^{-d}=\vartheta_n^{-d}.
$$
\vspace{3ex}

\begin{center}
{\bf 6. Conclusion}
\end{center}
\vspace{1ex}

The main result of the present work is revealing of fractal properties of the large-scale structure of the Universe which are described by the laws (2), (8), (9), (10) and (14).

Note that de Vaucouleurs tried to find a size distribution law of galaxies using the first Zwicky's and Abell's catalogues in 1971 \cite{ref5}. He concluded that galaxy distribution differs from a random Poisson one, but he couldn't find a law analogous to (14) due to insufficient volumes of the catalogues. The first effective application of the wavelet analysis on a plane for revealing of clusters and filaments on a small area of the celestial sphere is demonstrated in paper \cite{ref6} and estimated angular size values of 33 structures with angular sizes from $20''$ to $11^{\circ}$ are presented. Using these sizes one can ascertain that the law (14) is satisfied for filaments with fractal dimension $d\approx1{.}16$.

The fractality of the size distribution of quasar groups may be accounted for in framework of the fractal cosmological model described in paper \cite{ref26}. Einstein's equations for the space-time metric are shown to have a class of solutions in which the metrics are related to each other by a simple transformation: a transition from one metric to another is just a multiplication by a constant factor (scale transformation or scaling).

Perhaps, in the epoch of quasars, the large-scale structure was composed of noninteracting spatial regions with metrics related to each other by a scale transformation. Large quasar groups discovered in this paper mark these spatial regions. In this case, clumps are physically self-similar. Sizes of two groups are related by a scale transformation: $r_i=qr_{i-1}$. The whole set of sizes forms a geometric progression of type (15). In this case, a number of groups and their size are related by the correlation (14).

The fractal dimension value $d\approx 2{.}08$ for the large quasar group distribution in sizes is compared to that of polygonal path of Brownian particle (length distribution of segments). This analogy indicates that initial density perturbations from which large quasar groups arise, apparently, had a thermal spectrum.


\vspace{3ex}

\begin{thebibliography}{99}
\bibitem{ref1} Giovanelli R., Myers S.T., Roth J., Haynes M.P. A 21 CM survey of the Pisces-Perseus supercluster. II - The declination zone +21.5 to +27.5 degrees. // Astron.J., 1986, v.92, p.250-274.
\bibitem{ref2} Zel'dovich Ya.B., Einasto J., Shandarin S.F. Giant voids in the universe. // Nature, 1982, v.300, p.407-413.
\bibitem{ref3} de Lapparent V., Geller M.J., Huchra J.P. A slice of the universe. // Astrophys.J., 1986, v.302, p.L1-L5.
\bibitem{ref4} Slezak E., de Lapparent V., Bijaoui A. Objective detection of voids and high-density structures in the first CfA redshift survey slice. // Astrophys.J.,  1993, v.409, p.517-529.
\bibitem{ref5} de Vaucouleurs G. The Large-Scale Distribution of Galaxies and Clusters of Galaxies. // Publication of the Astronomical society of the PACIFIC, 1971, v.83, p.113-143.
\bibitem{ref6} Escalera E., MacGillivray H.T. Topology in galaxy distribution: method for a multi-scale analysis. A use of the wavelet transform. // Astron.\&Astroph., 1995, v.298, p.1-21.
\bibitem{ref7} Frith W.J., Outram P.J., Shanks T. The 2MASS galaxy angular power spectrum: Probing the galaxy distribution to Gigaparsec scales. // arXiv:astro-ph/0507215v2.
\bibitem{ref8} Sylos Labini F., Vasilyev N.L., Baryshev Y.V., L\'opez-Corredoira M.  Absence of anti-correlations and of baryon acoustic oscillations in the galaxy correlation function from the Sloan Digital Sky Survey data release 7. Astron.\&Astroph., 2009, v.505, p.981-990.
\bibitem{ref9} Choi Y.-Y., Park C., Kim J., Gott III R., Weinberg D.H., Vogeley M.S., Kim S.S.  Galaxy Clustering Topology in the Sloan Digital Sky Survey Main Galaxy Sample: a Test for Galaxy Formation Models. //  arXiv:1005.0256v2 [astro-ph.CO].
\bibitem{ref10} Haberzettl L., Williger G.M., Lauroesch J.T., et al. The Clowes-Campusano Large Quasar Group Survey: I. GALEX selected sample of LBGs at $z\sim 1$. // arXiv:0906.4058v1 [astro-ph.CO].
\bibitem{ref11} Pilipenko S.V. On the spatial quasar distribution. // Astronomicheskij Zhurnal, 2007, v.84, N.10, p.910-919.
\bibitem{ref12} Clowes R. G., Campusano L. E., Graham M. J., Sochting I. K. Two close Large Quasar Groups of size $\sim 350$ Mpc at $z\sim 1.2$ . // arXiv:1108.6221v1 [astro-ph.CO].
\bibitem{ref13} Granett B. R., Neyrinck M. C., Szapudi I.  An Imprint of Super-Structures on the Microwave Background due to the Integrated Sachs-Wolfe Effect. // arXiv:0805.3695v2 [astro-ph].
\bibitem{ref14} Ho S., Hirata C.M., Padmanabhan N., Seljak U., Bahcall N. Correlation of CMB with large-scale structure: I. ISW Tomography and Cosmological Implication. // arXiv:0801.0642v2 [astro-ph].
\bibitem{ref15} Shneider D.P., Richards G.T., Hall P.B., Strauss M.A. et al.  The Sloan Digital Sky Survey Quasar Catalog V. Seventh Data Release. // arXiv:1004.1167v1 [astro-ph.CO].
\bibitem{ref16} http://lambda.gsfc.nasa.gov/
\bibitem{ref17} Smith J.A., Tucker D.L., Kent S., Richmond M.W. et al. The u'g'r'i'z' Standart Star Network. // arXiv:astro-ph/0201143v2.
\bibitem{ref18} Boyle B.J., Shanks T., Peterson B.A.  The evolution of optically selected QSOs. II // MNRAS, 1988, v.235, p.935-948.
\bibitem{ref19} Agapov A.A., Rozgacheva I.K. Observational fractal properties of the quasar distribution according to SDSS catalogue. // Nelineyniy mir, 2011, v.9, N.6, p. 384-390.
\bibitem{ref20} Rozgacheva I.K.,  Agapov A.A. Fractal properties of SDSS quasars. // arXiv:astro-ph.1101.4280.
\bibitem{ref21} Baryshev Yu., Teerikorpi P. The fractal analysis of the large scale galaxy distribution. Bull. Spec. Astrophys. Obs., 2006, V.59, P.92-154.
\bibitem{ref22} Jones B.J.T, Martines V.J., Saar A., Trimble V. Scaling Laws in the Distribution of Galaxies. // arXiv: astro-ph/0406086.
\bibitem{ref23} Peebles P.J.E., Hauser M.G. Statistical analysis of catalogs of extragalactic objects. III. The Shane-Wirtanen and Zwicky catalogs. // Astrophys.J.Suppl.Ser., 1974, v.28, �253, p.19-36.
\bibitem{ref24} Peebles P.J.E., The large-scale structure of the Universe. Moscow: Mir, 1983.
\bibitem{ref25} Bond J.R., Efstathiou G. The statistics of cosmic background radiation fluctuations. // MNRAS, 1987, v.226, p.655-687.
\bibitem{ref26} Rozgacheva I.K., Agapov A.A. The fractal cosmological model. // arXiv:astro-ph.1103.0552.
\bibitem{ref27} Jackiw R. Introducing Scale Symmetry. Phys. Today, 1972, V.25 (1), p.23.
\bibitem{ref28} Canuto V., Adams P.J., Hsieh S.-H., Tsiang E. Scale-Covariant Theory of Gravitation and Astrophysical Application. // Physical Rev.D, 1977, v.16, N.6, p.1643-1663.
\bibitem{ref29} Antoine J.-P., Demanet L., Jacques L., Vandergheynst P. Wavelets on the Sphere: Implementation and Approximations. // Applied and Computational Harmonic Analysis, 2002, v.13, � 3, p.177-200.
\bibitem{ref30} Feder J. Fractals. Moscow: Mir, 1991.
\end{thebibliography}
\end{document}